\documentclass[structabstract]{aa}  
\usepackage{graphicx}
\usepackage{txfonts}
\begin{document}
   \title{Multiple and changing cycles of active stars}

   \subtitle{I. Methods of analysis and application to the solar cycles}

   \author{Z. Koll\'ath
          \inst{1}
          \and
          K. Ol\'ah\inst{1}
          }

\offprints{Z. Koll\'ath}

   \institute{Konkoly Observatory of the Hungarian Academy of Sciences,
              P.O.Box 67, H-1525 Hungary,
              \email{kollath@konkoly.hu}
   }

   \date{Received ; accepted}

\authorrunning{Koll\'ath \& Ol\'ah}
\titlerunning{Changing cycles -- method}

  \abstract
   {}
{Long-term observational data have information on the magnetic 
cycles of active stars and that of the Sun. The changes in the activity of our 
central star have basic effects on Earth, like variations in the global 
climate. Therefore understanding the nature of these variations is 
extremely important. }
{The observed variations related to magnetic activity cannot be treated as stationary periodic variations, therefore methods like Fourier transform or different versions of periodograms give only partial information on the nature of the light variability. We demonstrate that time-frequency distributions provide useful tools for analysing the observations of active stars.}
{With test data we demonstrate that the observational noise has practically no effect on the determination in the the long-term changes of time-series observations of active stars. The rotational signal may modify the determined cycles, therefore it is advisable to remove it from the data. Wavelets are less powerful in recovering complex long-term changes than other distributions which are discussed. Applying our technique to the sunspot data we find a complicated, multi-scale evolution in the solar activity.
}
{}


   \maketitle
%

\section{Introduction}

It is well known that solar cycles cannot be represented by single periodic variations. Besides the 11 year quasi-periodicity (Schwabe \cite{schwabe}) of solar activity lots of different rhythms were discovered or at least suggested by different studies (see e.g. Frick et al. \cite{frick}, Forg\'acs-Dajka \& Borkovits \cite{emese}). One remarkable and well-studied periodicity of the solar variation was found by Gleissberg (\cite{gleissberg}) about 70 years ago. He described this variability as a "long-periodic fluctuation of the sun-spot numbers, the period being about seven or eight sun-spot cycles". Frick et al. (\cite{frick}) derived the length of the Gleissberg cycle as $\approx$100 years from wavelet analysis of a 384 years long dataset (Hoyt et al. \cite{hoyt}). 
The non-stationarity in the long-term variation of active stars has already been found in active stars as well (Ol\'ah et al. \cite{cycles3}).

 Time-frequency representation of a non-stationary signal yields information about characteristics of the dataset in the time-frequency plane. Then transient events or frequency/amplitude changes can be monitored easily with these tools. In the literature, a frequently used method is the wavelet-map and some other techniques based on wavelets. However, wavelet is only one of the several time-frequency representations, and other methods may give better results than wavelets. The key problem is the balance between temporal and frequency resolution. The frequency-time uncertainty principle gives strict constraints for the time duration ($\Delta t$) and the bandwidth ($\Delta \omega = 2\pi \Delta f$) of a signal: $\Delta t * \Delta \omega \ge 1/2$. These measures of broadness are determined by  the standard deviation of frequency and time, and the uncertainty principle with these quantities applies for any time-frequency distribution. The linear transforms (such as wavelet or windowed Fourier-transform) have a limit for joint resolution of time and frequency for local structures: for example in general it is not possible to determine the frequency precisely at high temporal resolution. Bilinear  time-frequency transformation, called kernel method (Cohen \cite{cohen}) can break the above principle locally, but its cost is paid by artifacts at other frequencies. 

While wavelets and other methods based on Fourier transforms are exclusively used to  analyse stellar activity, more sophisticated methods have been available for decades: the Wigner distribution (Wigner, \cite{wigner}) and its extensions by Cohen (\cite{cohen}). These methods have also been proven to be useful in the  analyses of variable star data of chaotic origin or non-stationary behaviour (see Buchler et al. \cite{buchler1}, Koll\'ath \& Buchler \cite{kollath1}). In this paper we present a method which can be used to study changes of stellar activity cycles in time using data with yearly gaps and with short-term signals (rotational modulation).

As example we reanalyzed the available sunspot and radio records with modern time-frequency methods to give a complete picture of the beating of the Sun in the  period range from a few years to 200 years. In the companion paper (Ol\'ah et al. \cite{paper2}) we apply the method described in the present paper to 20 active stars, and discuss the time variability of their cycles.


\section{Time-frequency distributions}

A simple but still powerful method is the 'short-term Fourier transform' (STFT hereafter). In this case the dataset is multiplied by a Gaussian window centred around a given epoch $t_0$. If this windowed part of the signal is then Fourier transformed, the frequency spectrum of the data around the time $t_0$ can be generated. By shifting the centre of the Gaussian window in time, a sequence of Fourier spectra is obtained, which gives the time-frequency representation of the data. If the sampling of $t_0$ is dense enough, then a two-dimensional map of the energy distribution on the time -frequency plane can be plotted. 

\begin{figure}[ht!]
   \centering
   \includegraphics[width=8.5cm]{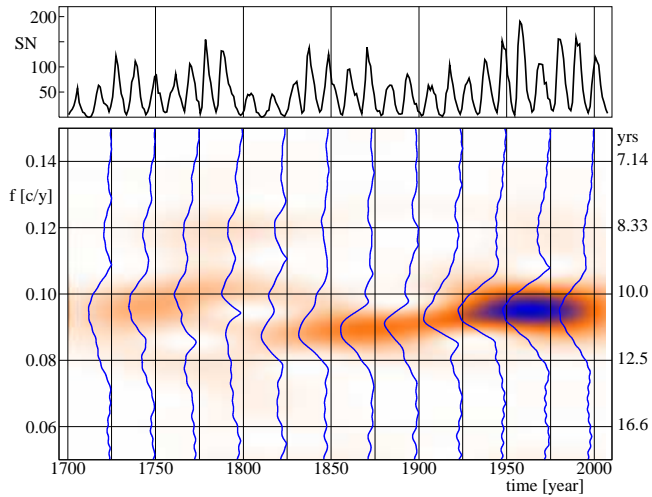}
   \caption{Short-term Fourier transform (bottom) of the sunspot 
number time series (top).
            Fourier-transforms of the subsets of the dataset are also  
            plotted. The colour scale of relative variations in the 
            distributions is presented in Fig. 7.}
              \label{Fig1}
    \end{figure}

We demonstrate this method on the yearly sunspot numbers -- a dataset which is well known and  analysed by several authors. Fig.~\ref{Fig1} displays the STFT of the solar data. On the time-frequency distribution map individual windowed Fourier spectra are also plotted with thick lines. By changing the width of the Gaussian window, the balance between time and frequency resolution can be modified. All the locations on the map, of course, satisfy the uncertainty principle $\Delta t * \Delta \omega \ge 1/2$, and the ratio of time and frequency resolution is constant on the map.

While in this demonstration we used a time centred Gaussian window to plot the curves of Fourier transforms, in the implementation of STFT we use windowing in the Fourier space:
\begin{equation}
 T(t,f) = 2 \int_0^{+\infty} F(\nu) \exp  
 \Big(  -{1\over 2} {(\nu-f)^2 \over \sigma^2(f)} \Big) 
 \exp (i2\pi t \nu) d\nu ,
\end{equation}
where $F(\nu)$ denotes the Fourier transform of the signal. Note, that the inverse Fourier transform is calculated only for the positive frequencies. This method has several advantages. At low frequencies it reduces the distortion due to the components at negative frequencies, moreover when $\sigma$ is large, the transform reduces to the analytic function (see Buchler and Koll\'ath \cite{buchler2}). Depending on the functional form of $\sigma(f)$, Eq.~1 reduces to STFT or wavelet. With $\sigma=f_0/\alpha$, where $f_0$ is a constant frequency, the equivalent temporal window function of STFT is given by 
\begin{equation}
 h(\tau) = \exp \Big( {-\tau^2 f_0^2 \over 2 \alpha^2} \Big) 
\end{equation} 
Usually it is a good choice to set $f_0$ to the maximum or the central frequency of the studied period domain. Then $\alpha \sim 1$ provides a naturally well balanced temporal and frequency resolution. It can be seen from Eq. 2 that the half width of the time centered windowing function is $\alpha/f_0$. Then, if $\sigma$ depends on frequency, the width of the window also varies with frequency. Especially 
if $\sigma(f)= f/\alpha$ then $T(t,f)$ is equal to wavelet, with the Morlet kernel, since the width of the temporal windowing function decreases with $1/f$.

\begin{figure}
   \centering
   \includegraphics[width=8.0cm]{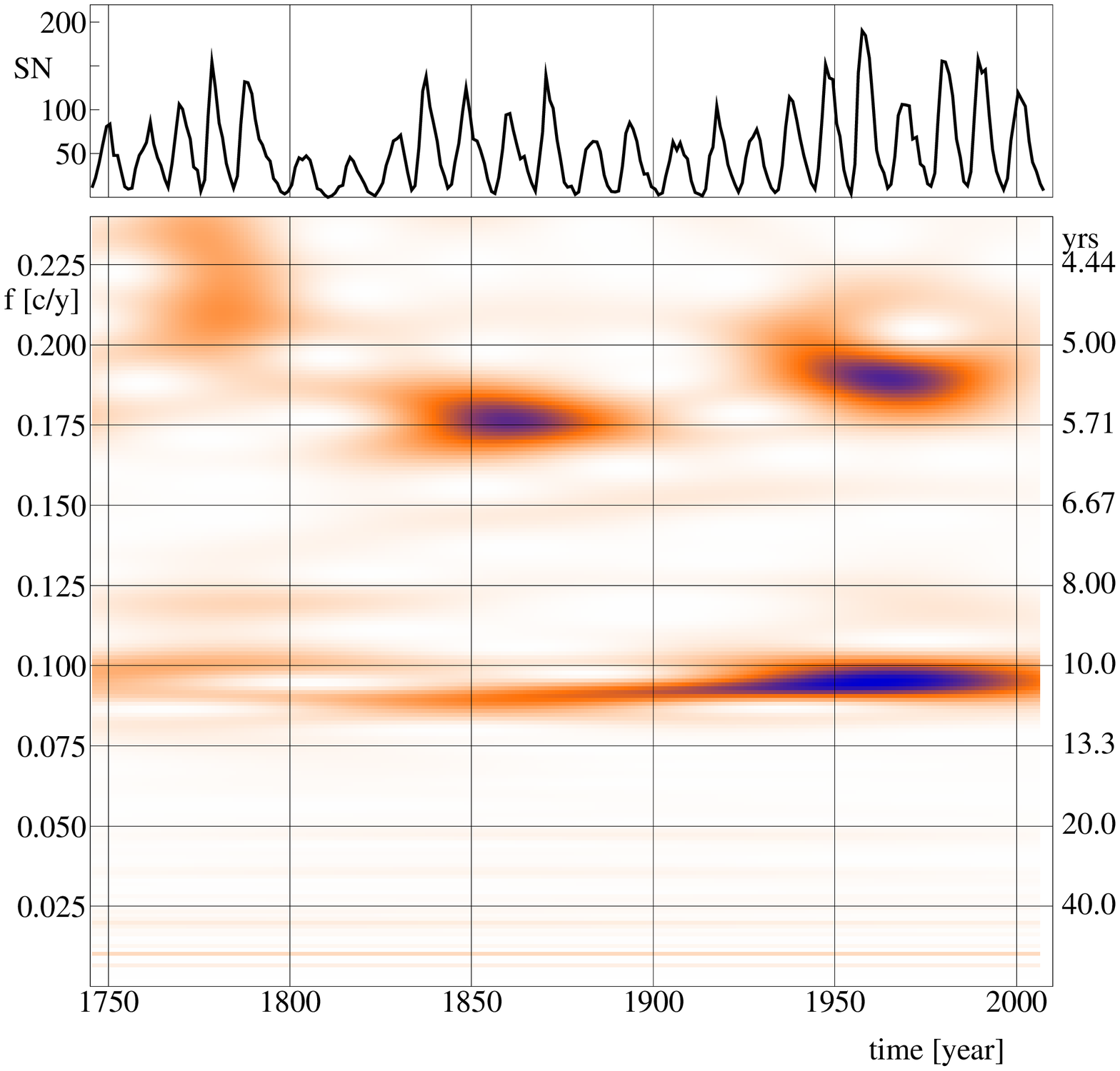}\vspace*{5mm}
   \includegraphics[width=8.0cm]{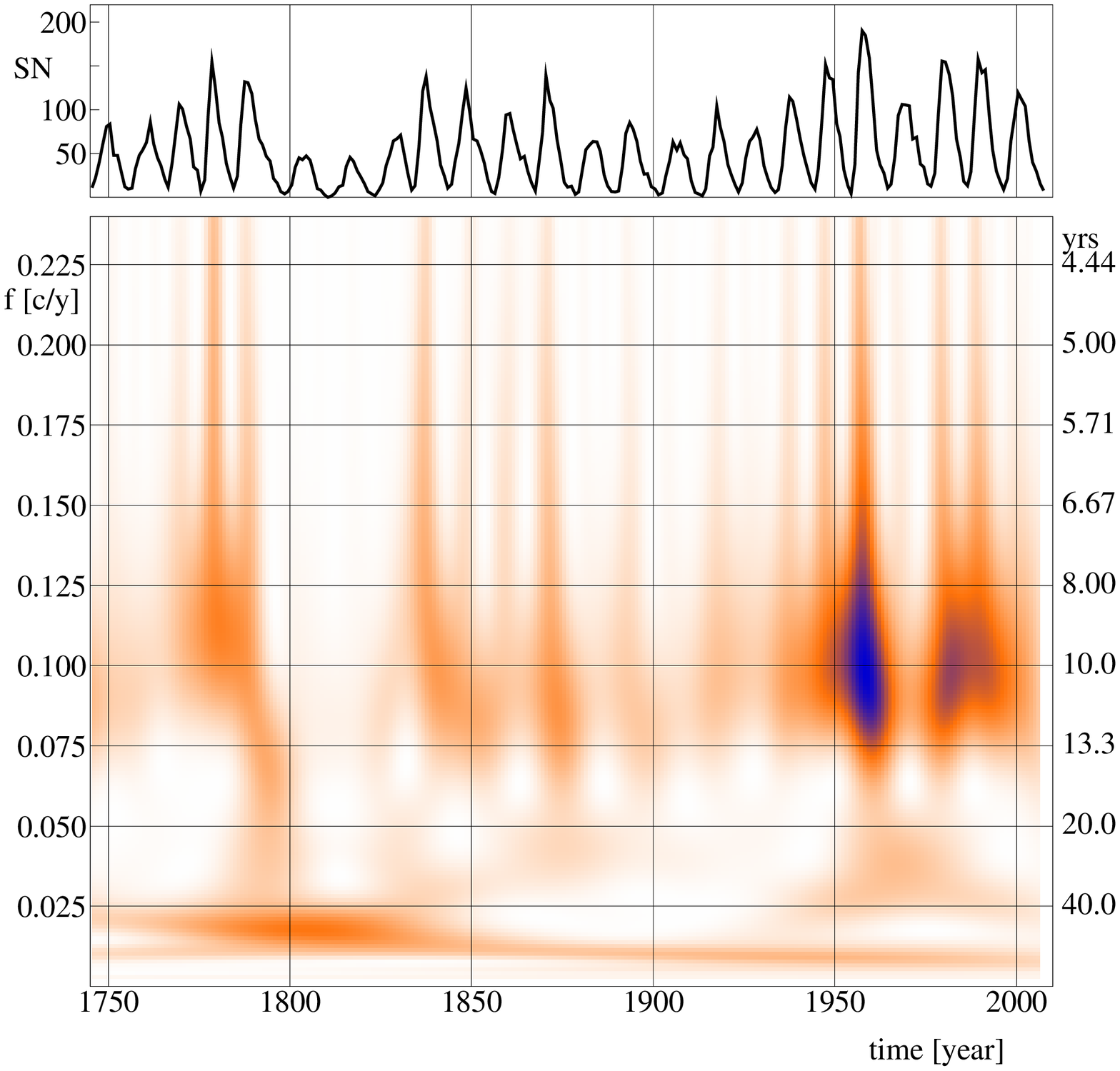}\vspace*{5mm}
   \includegraphics[width=8.0cm]{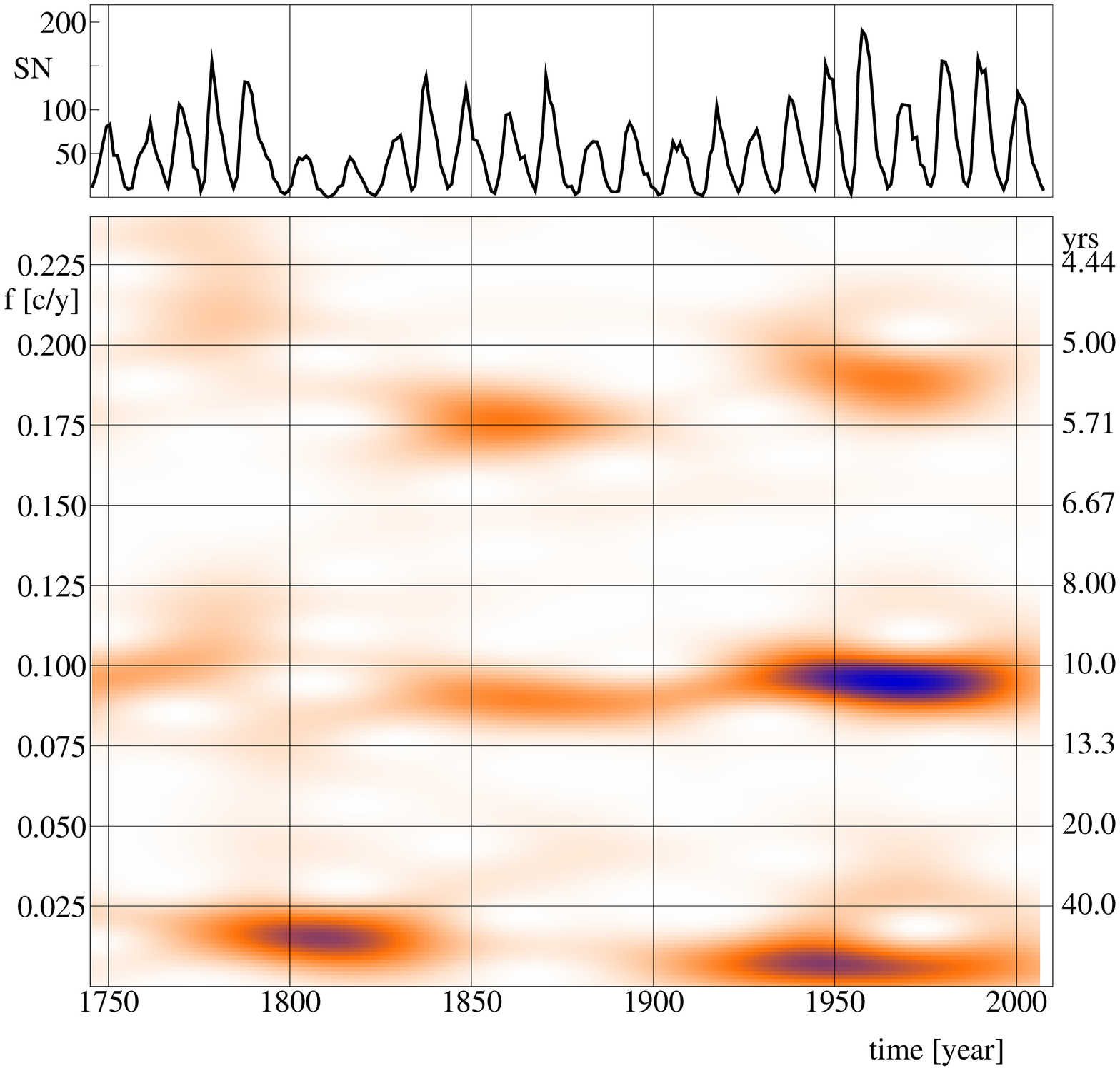}
   \caption{Wavelet (top and middle) and short-term Fourier transform (STFT) distribution (bottom) of the yearly sunspot numbers.
            }
              \label{Fig2}
\end{figure}

Note that in the standard Morlet wavelet $\alpha$ is fixed to $1$. The introduction of a free parameter ($\alpha$) which controls the balance between frequency and temporal resolution has several advantages as already described by Koll\'ath \& Szeidl (\cite{kollath2}) and Soon, Frick \& Baliunas (\cite{soon}).

\begin{figure*}
   \centering
   \includegraphics[width=7.6cm]{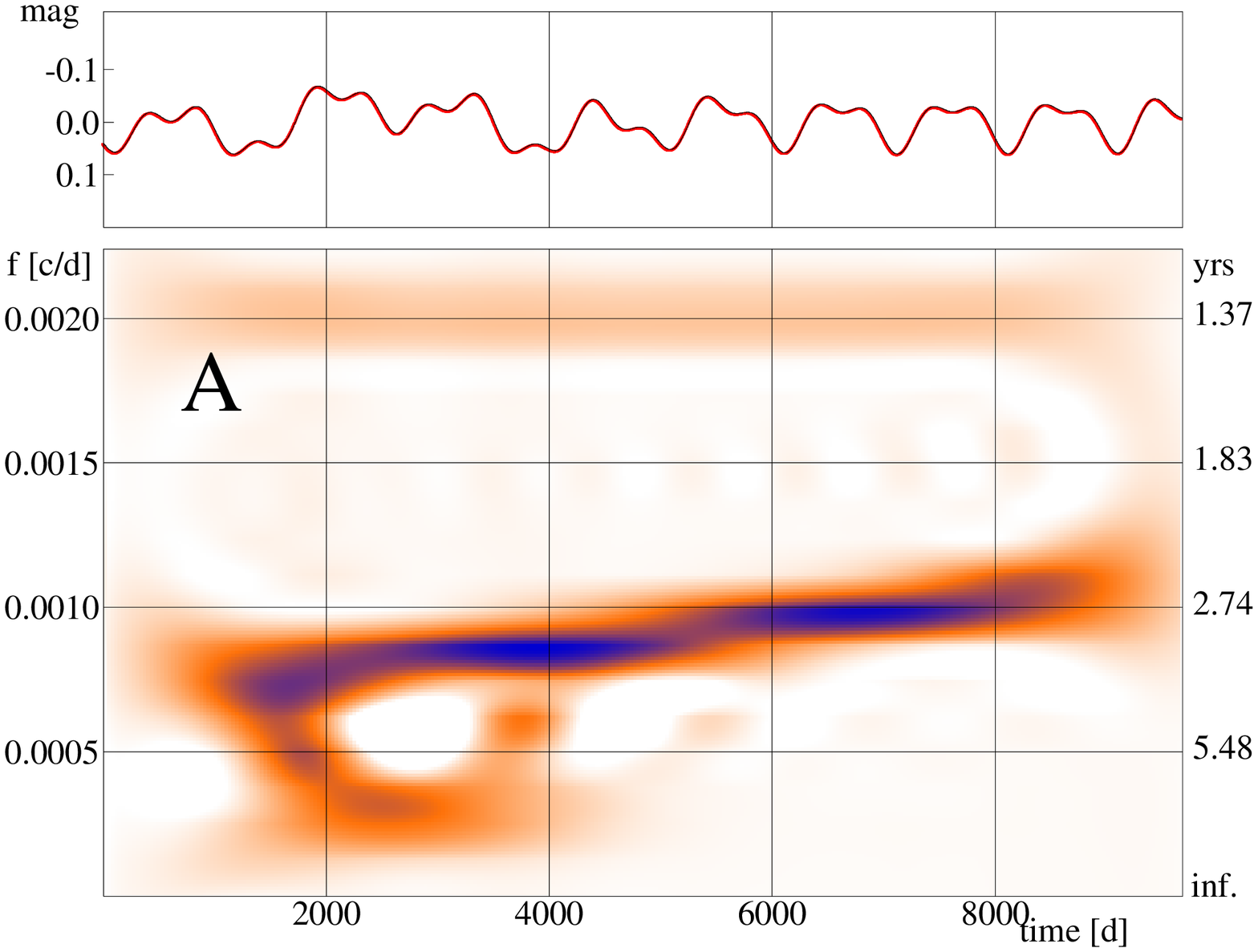}\hspace*{5mm}
   \includegraphics[width=7.6cm]{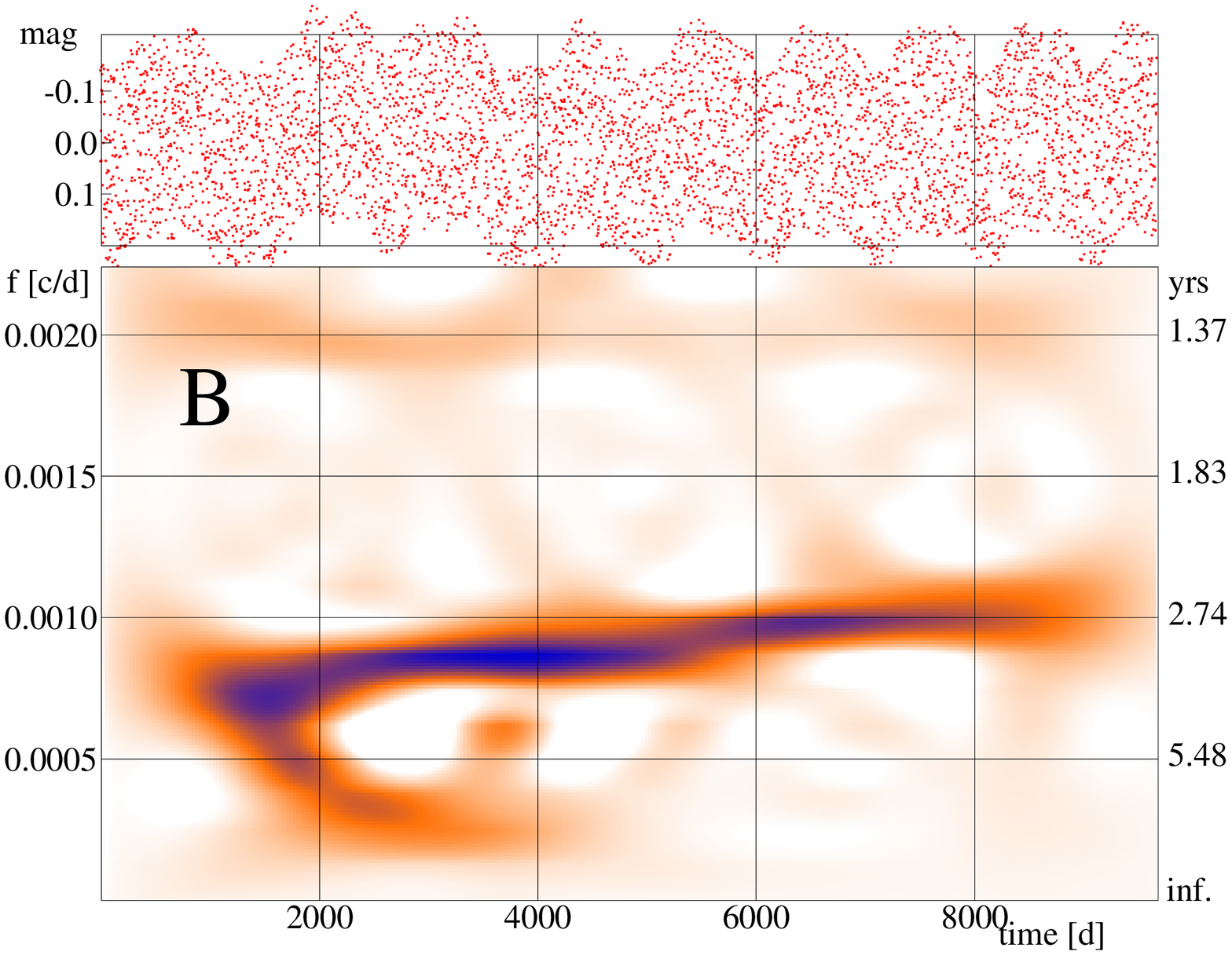}\vspace*{3mm}
   \includegraphics[width=7.6cm]{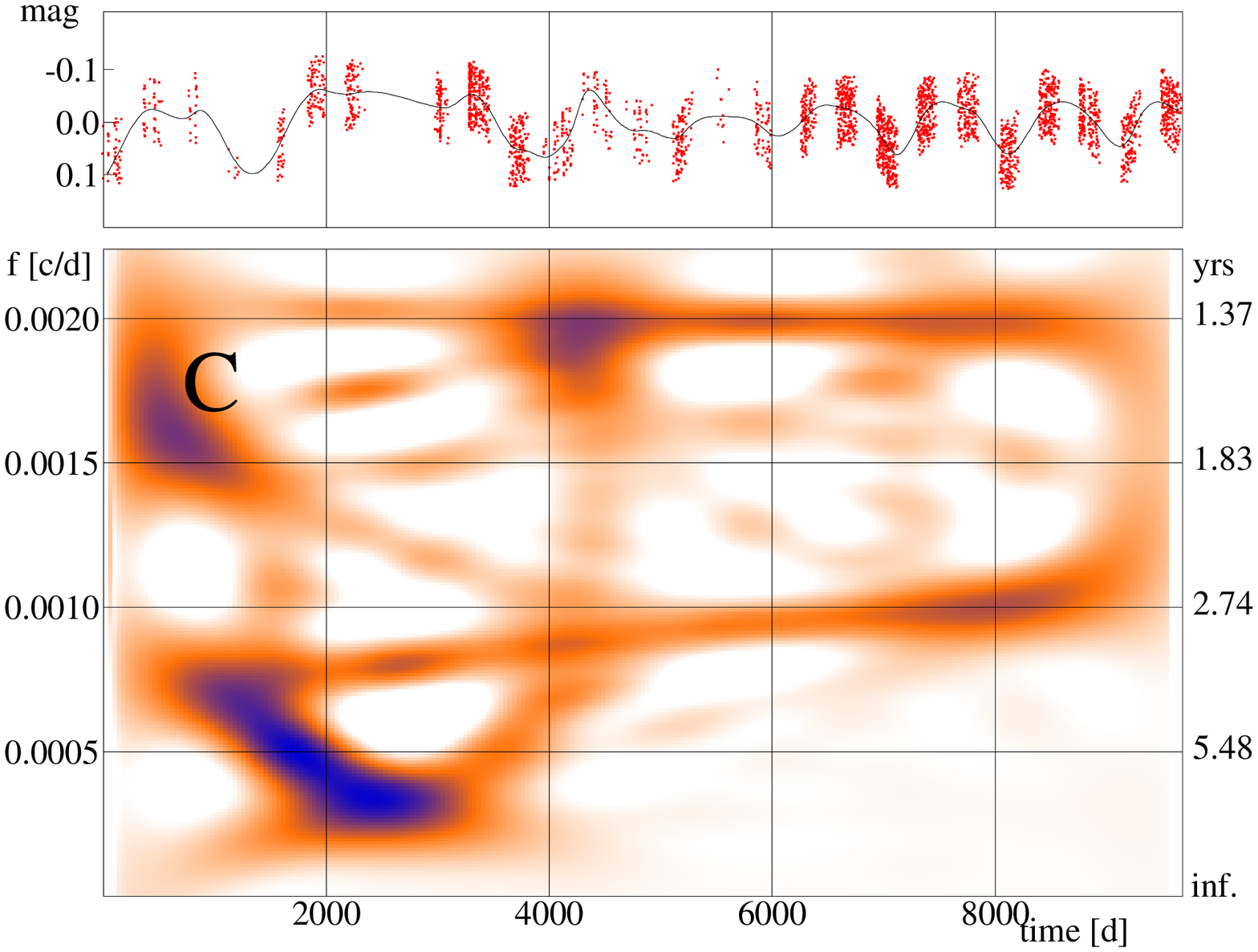}\hspace*{5mm}
   \includegraphics[width=7.6cm]{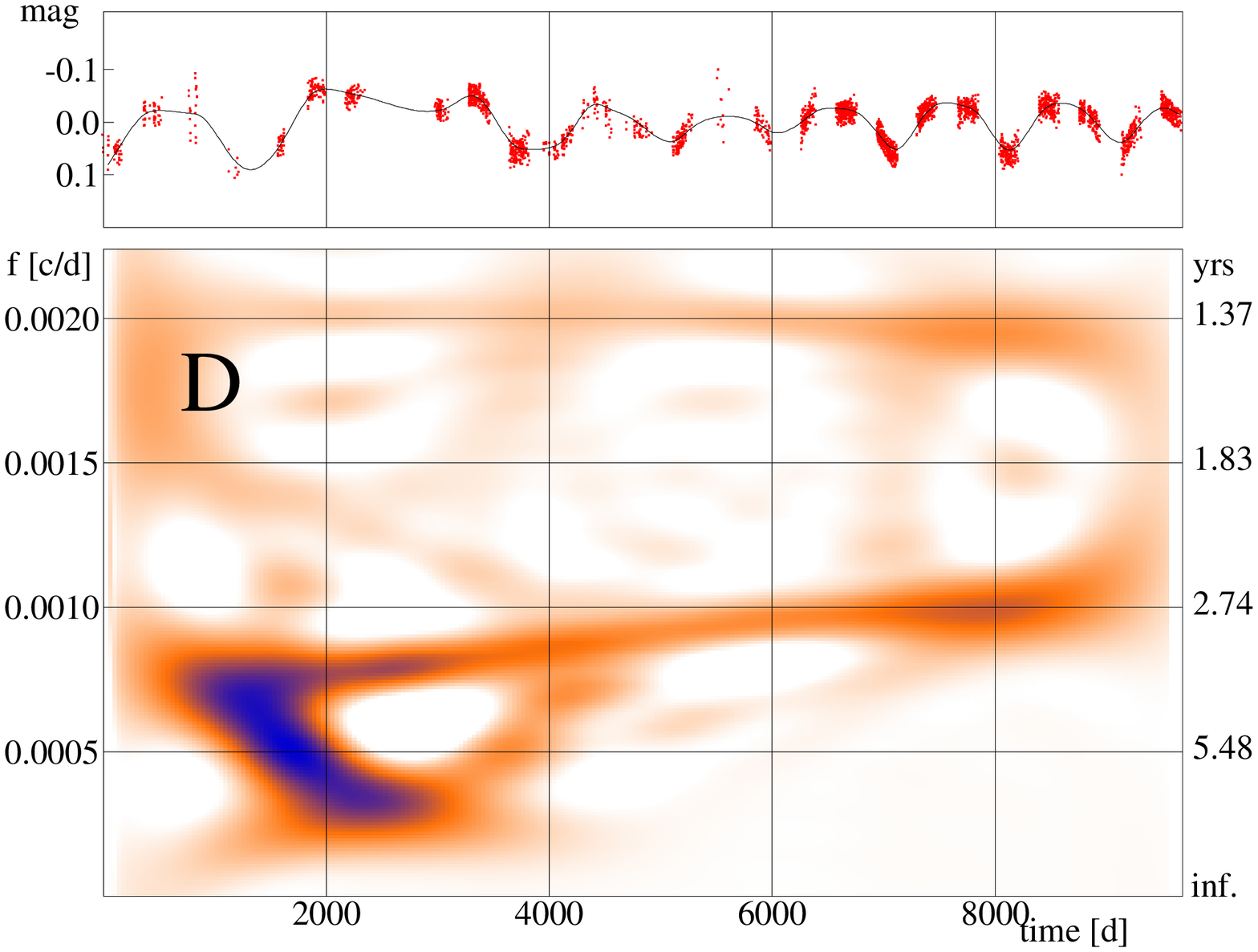}\vspace*{3mm}
   \includegraphics[width=7.6cm]{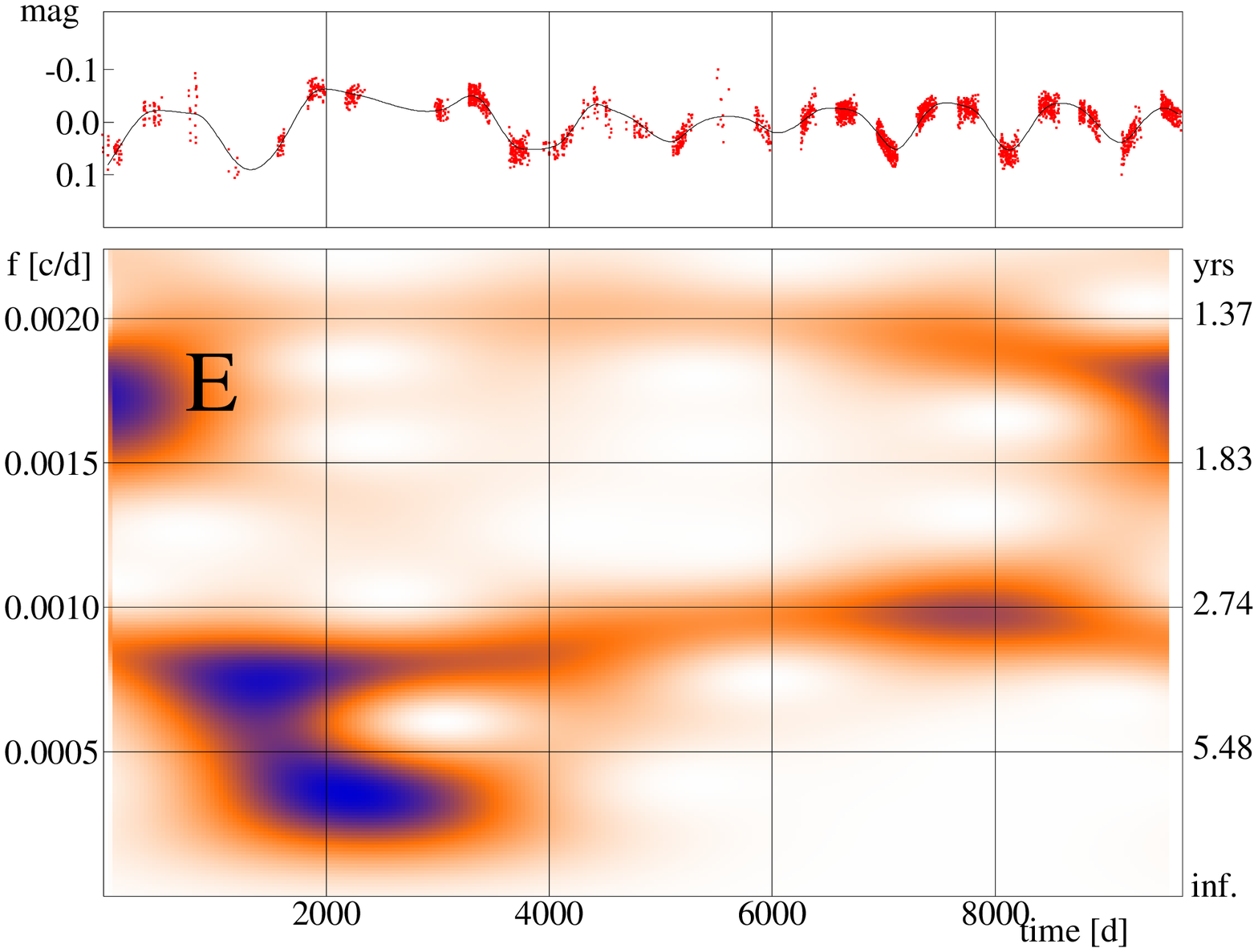}\hspace*{5mm}
   \includegraphics[width=7.6cm]{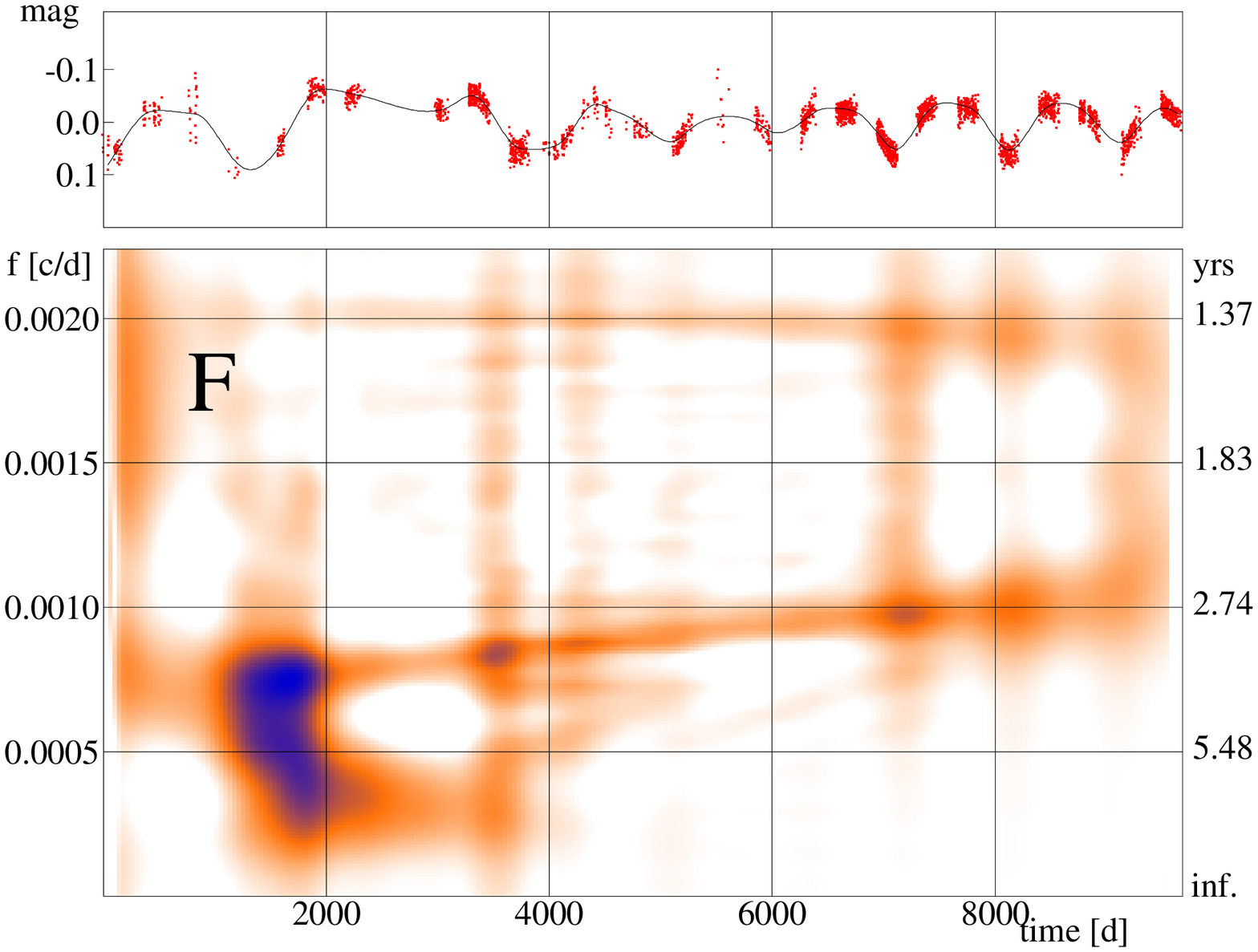}\vspace*{3mm}
   \caption{Time-frequency distribution of test data. {\it Panel A}: PWD of signal 'I'; {\it panel B}: PWD of signal 'I' with white noise; {\it panel C}: PWD of signal 'II' with averaging and spline interpolation; {\it panel D}: PWD of signal 'II' after Fourier filtering of the rotational period in the independent observational seasons and then averaging and spline smoothing; {\it panel E}: STFT of the same signal as in panel {\it D}; {\it panel F}: CWD of the same signal as in panel {\it D}.
           }
              \label{Fig3}
\end{figure*}

As can be seen from the above definitions, wavelet with Morlet-kernel is very similar to STFT, however, for wavelets the balance between temporal and frequency resolution depends on the frequency. This feature is demonstrated in the top panels of Fig.~\ref{Fig2}. The parameter ($f_0$) is defined in such a way that for a given $\alpha$ the wavelet and STFT have the same frequency/time resolution at the frequency $f_0$. The top and middle panels of Fig.~\ref{Fig2} show wavelets where the lower frequency part (top) and the high frequency part (middle) was matched to the STFT, respectively. Resolution in time increases with frequency, since the same number of cycles is covered in a shorter time-base, at higher frequencies. The disadvantage of this representation is that harmonic structures, which display the same characteristics in STFT, are distorted. Fig.~\ref{Fig2} thus demonstrates that while the time-frequency distribution in the lower panel clearly show the synchronised variation of the $\approx$11 year cycle with its double frequency harmonic, the wavelet destroys this parallel structure. 

Note that we used a different scaling of the amplitudes at different frequencies. In time-frequency distributions the lower amplitude parts are easily washed out by the contributions from power of other components. Far from our topics but similar in mathematical tools, in speech analysis it is well known that by the application of a pre-emphasis filter the obscured parts of the frequency distribution can be recovered. A multilevel pre-emphasis Fourier filter is applied to visualise the whole studied period range equally. For the solar data, relative to the highest amplitude $\approx$11 year periodicity (0.07 -- 0.14~$c/y$) the signal in different frequency ranges were amplified by the following factors ($R_e$): $R_e=1.75$ for $0.0 < f < 0.05$, and $R_e=3.5$ above $f=0.16$ ($f$ is given in $c/y$). In the transition regions a smooth change in $R_e$ is used. With this pre-emphasis filtering  there is a factor of 3.5 difference in the amplitudes to get similar grey levels around 11 and 5.5 years periods.

The optimal balance of temporal and frequency resolution depends on the signal itself. A question arises whether it is possible to construct a distribution, where the resolution in both coordinates is optimal, but some other disadvantage is allowed as its cost. The answer is a more general group of methods for signal processing introduced by Cohen (see e.g. Cohen \cite{cohen}). The generalised time-frequency distribution is given by
\begin{eqnarray}
 C(t,f) = {1\over 2\pi} \int\int\int
 exp(-i\xi t - 2\pi\, i\tau\, f - i\xi\theta) \nonumber \\
 \Phi(\xi,\tau) s^*(t-\tau/2) s(t+\tau/2)
 d\theta d\tau d\xi,
\end{eqnarray}
where $s(t)$ is the  analysed time series and $\Phi(\xi,\tau)$ is the kernel of the distribution that determines the specific properties of the distribution. The simplest time-frequency distribution is the Wigner-Ville transformation with $\Phi(\xi,\tau)=1$. This distribution works well only for special datasets, otherwise the resulting map is heavily contaminated by cross terms of the different components. A bi-Gaussian kernel provides a distribution which, in the limit of the kernel width, is equivalent to STFT. This so-called pseudo-Wigner distribution (PWD hereafter) is widely used in sound processing. For comparison purposes we also use the Choi-Williams distribution (CWD) (see Choi and Williams \cite{cwd}), which applies an exponential kernel. 

The top panel of Fig.~\ref{Fig6} shows the PWD of the yearly sunspot numbers. It can be seen that the resolution in both coordinates is improved compared to the result with STFT displayed in Fig.~\ref{Fig2} (bottom panel), however, there are artificial features visible on the PWD distribution which have no real meaning in the observed signal. The reason for these cross terms is the non-linearity of the method: e.g., if two sinusoidal waves with different frequencies coexist in the signal then false power arises between the two ridges representing the cycles.

\begin{figure}
   \centering
   \includegraphics[width=8.5cm]{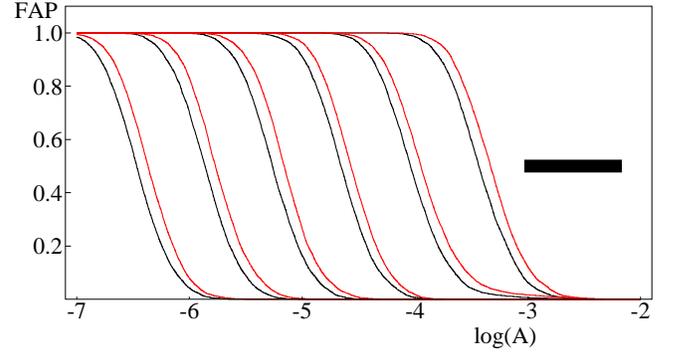}

   \caption{False alarm probability that a structure appears in STFT (black) and CWD (grey) due to observational noise. The amplitude of the added Gaussian noise (from left to right): 0.0025, 0.005, 0.01, 0.02, 0.04 and 0.08. The black rectangle represents the amplitudes of the visible real structures in Fig.~\ref{Fig3}.
            }
              \label{Fig4}
\end{figure}

\section{Preprocessing observational data}

Stellar observations in reality have lots of deficiency, like observational noise and imperfect sampling. Then time-frequency methods cannot be applied directly to the observational data, the analyses should be preceded by a sequence of preprocessing steps.

In the case of active stars the situation is even more complicated: different timescales of variability coexist which interfere with the sampling in different ways. The rotational periods fall into the range of days to weeks, very close to the order of the sampling time. On the contrary, long term changes have a characteristic time above a year, so averaging and resampling on this level is needed to obtain information on these components. Thus, the preprocessing of the data requires extreme care.

To avoid false averages due to the under-sampling of the rotational component, it is advisable to whiten all observations with the rotational period. Because the rotational component is modulated due to e.g. the differential rotation, prewhitening cannot be performed simultaneously for the whole dataset. 

The main problem is that the light-curves are unevenly sampled, and in addition, they contain seasonal gaps. Although there exist methods to calculate e.g. wavelets of this kind of data directly (like the adaptive wavelet), according to our previous experiments we prefer to interpolate the data when it is possible. When the sampling is denser than the dominant periodicities in the data, then a simple moving averaging can provide a continuous signal. Next, the averaged data can be smoothed with a cubic smoothing spline (Reinsch \cite{reinsch}). This operation provides a continuous function $S(t)$ based on the discrete time series ($\{t_i,x_i\}$ $i=1,n$) by minimising the integral of $S"(t)^2$, with the following constraint: 
\begin{equation}
{1\over n} \sum_{i=1}^n (x_i-S(t_i))^2 \le \sigma_s,
\label{eq_spline}
\end{equation}
where $\sigma_s$, the smoothing factor is a free parameter, which controls the smoothness of the spline. In theory its value should be chosen in the confidence interval corresponding to the left side of Eq. \ref{eq_spline} thus $\sigma_s$ should be of the same order as the standard deviation of the observational noise. In practice, we usually check manually the resulting spline smoothed curves to make a fine tuning of the averaging and spline smoothing. As a final step $S(t)$ is resampled to generate an equally spaced time series.

\section{Testing the methods}

\subsection{Artificial test data sets}

To test the methods for a general signal representing the cycles of active stars, we created artificial datasets. Signal 'I' consists of a periodic signal with a constant frequency of $0.002~c/d$, a component with linearly increasing frequency (chirp) with a modulation between $f=0.0007$ and $f=0.0011~c/d$ and a wave packet at $f=0.0003~c/d$ with a Gaussian amplitude modulation:$exp(-(t-2500)^2/2000^2)$). The amplitudes of the constant $f=0.002~c/d$ signal, the chirp and the wave packet are 0.02, 0.04 and 0.04 repectively. 

In addition to the components of signal 'I', signal 'II' has a higher frequency variation, demonstrating the rotational period of active stars, and a uniformly distributed noise with amplitude comparable to the rotational component. To test the preprocessing and the time-frequency methods together, the test signal 'II' was finalised by resampling the data with the time series of the observations of EI~Eri (see Paper II) The data were averaged with a time base of 90 days, then the averages were spline smoothed with $\sigma_s = 0.005$ and resampled with $\Delta t = 10$days.

\subsection{Observational noise and time-frequency distributions}

It is well known that in case of periodic or quasi-periodic signals the Fourier transform has a very good noise rejection property. The same is true for time-frequency distributions. If the observational noise and the signal are not correlated, then the method provides good result even in the case of large noise. A nice example of this nature of time-frequency distributions is the analysis of the semi-regular star V CVn (Buchler, Koll\'ath \& Cadmus \cite{buchler3}), in which two datasets are available for the same period of time. One of the records consists of visual magnitude estimates by amateur astronomers, while the other data are professional measurements by photoelectric photometry. It is clear that even the statistical properties of the noise sources are different in the two cases. Figure 2 in Buchler, Koll\'ath \& Cadmus (\cite{buchler3}) clearly demonstrates that the time-frequency plots are almost fully independent of the choice of the dataset, so the effect of observational noise is negligible for signals with good SNR and sampling.

We demonstrate this noise rejection property of time-frequency distributions on our first test time-series. The top panels of Fig.~\ref{Fig3}. display the PWD of signal 'I' (A) and the same signal modified with noise, with large relative temporal variance, added to it (B). The amplitude of the noise is seen on the top panel of the figure.  While the noise has practically no effect on the lower frequencies, the $f=0.002~c/d$ component is slightly distorted. The results with signal 'I' are presented only for PWD, but we note that the other time-frequency methods (STFT, wavelet, CWD) have the same properties.

We have tested the sunspot observations for the sensitivity to noise, too. White or Gaussian noise with the same amplitude as the maximum sunspot number does not significantly distorts the time-frequency plots. 

\begin{figure}
   \centering
   \includegraphics[width=8.5cm]{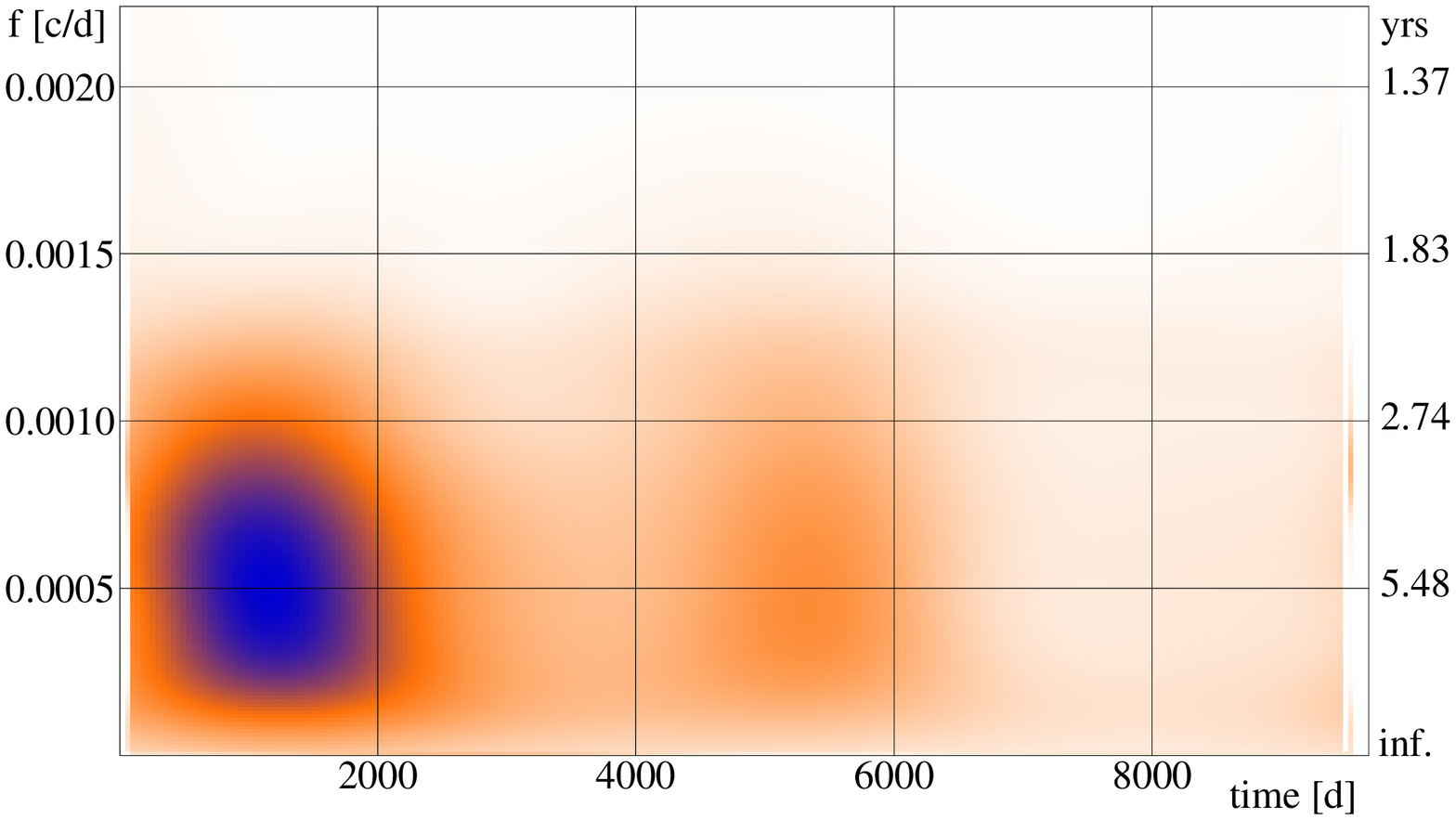}\vspace*{3mm}
   \includegraphics[width=8.5cm]{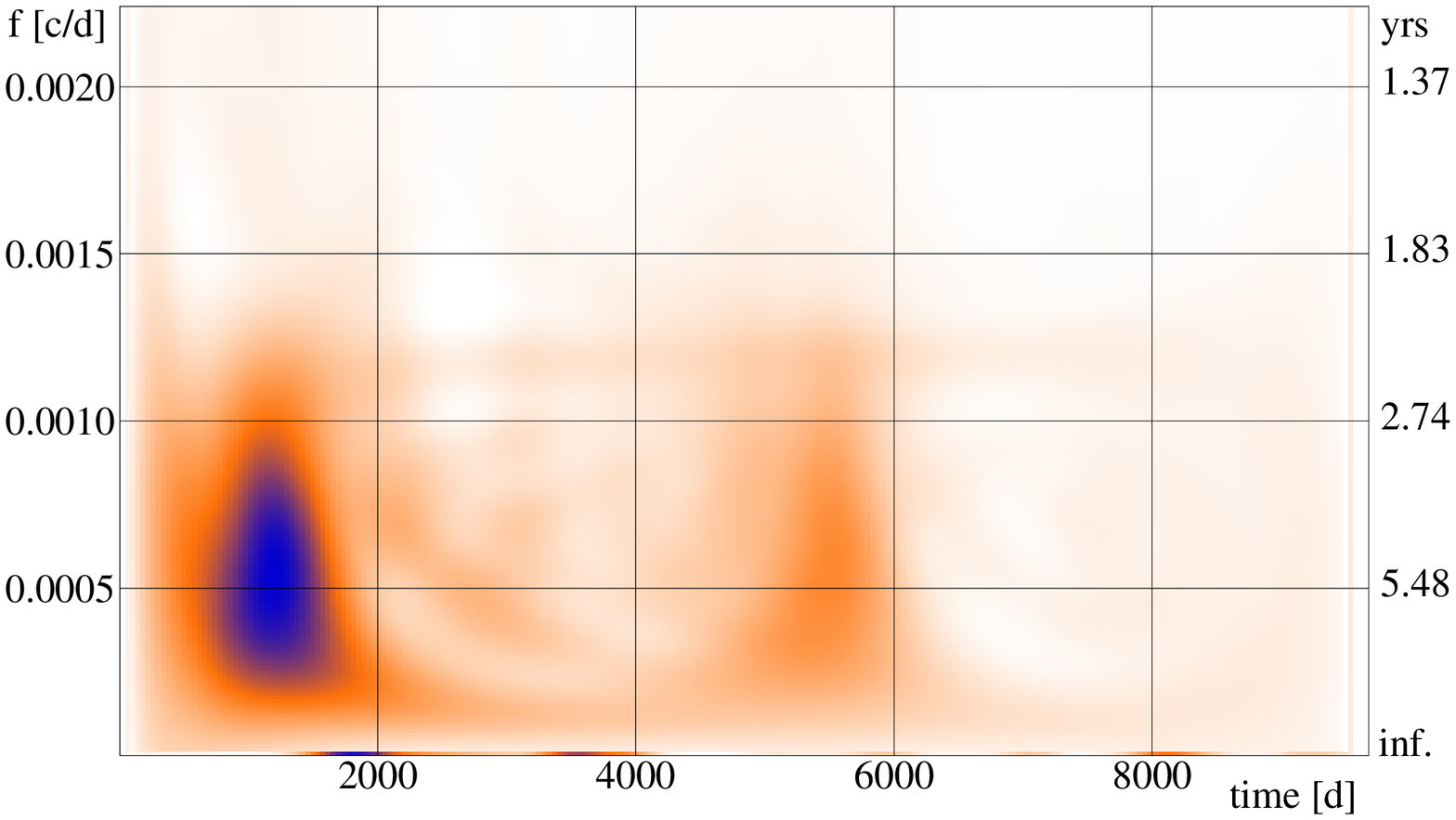}

   \caption{Time-frequency transfer function of the processing. Top: STFT, bottom: PWD}
              \label{Fig5}
\end{figure}

\subsection{Effect of uneven sampling and noise}

Because of the averaging and spline interpolation, the effect of observational noise is mixed with the effects of sampling, we performed our tests with ill-sampled data.  The resulting time-frequency distributions obtained from the gapped signal 'II' are displayed in the lower panels of Fig.~\ref{Fig3}, where, on panel C the rotational signal is included to, while on panels D, E and F it is filtered out from the data.
Averaging and smoothing spline interpolation introduce a low-pass filtering of the signal, which can be compensated by the pre-emphasis filtering. In these cases we selected $R_e=4.0$ for $f> 0.0015$. With these settings and processing, all the components above about 2 years of characteristic time are recovered well, especially in the case when the rotation period is filtered out. Traces of the $F=0.002$c/d ($P=1.37$yrs) component can be found, but it is strongly distorted. The reason for this distortion is, that due to the sampling, these periodicities cannot be exactly recovered. Except this deficiency, the main time scales of the ill-sampled data can be recovered by the methods described above. This result gives a warning that, because of the gapped nature of stellar data, activity cycles around 1-2 years cannot be easily, if at all, recovered.

The above conclusions are valid for all the time-frequency distributions we tested. We display the results for the filtered signal 'II' for three different transformations. The main components are recovered with all distributions. The cross terms are clearly visible in PWD and CWD (especially in CWD). Closely spaced frequencies, together with their cross terms tend to form loops (or 'bubbles'), like the ones visible in all PWD figures between $t \approx$ 2000 and 4000 days at low frequencies. These are well defined signatures of structures with two different frequencies, even short living ones, and their appearance remains mostly invariant against changes in the preprocessing (e.g. smoothing). STFT with our test data clearly displays the same double frequency structure, however, according to our experience, in some cases PWD is superior to STFT in this property.

To give a quantitative test on the combined effect of sampling and noise in a statistical sense, we performed Monte-Carlo simulation of the effect of observational noise on signal 'II'. We calculated $10^4$ realisations of the added Gaussian noise sequence with different amplitude (standard deviation). Then for each realisation we calculated the same procedure as for the original test data. The time-frequency distribution of the difference of the noisy and the noiseless smoothed data was obtained and the statistics of the amplitude of the largest feature (peak or ridge) was estimated. Fig.~\ref{Fig4} shows the False Alarm Probability $FAP(A)$, i.e., the probability that a structure with amplitude larger than $A$ appears in the transform due to observational noise. This test was performed with different standard deviations ($\sigma_n$) of the added Gaussian noise, and for all the time-frequency distributions we use. On the figure from left to right the curves belong to $\sigma_n$= 0.0025, 0.005, 0.01, 0.02, 0.04 and 0.08 mags, representing a wide range of observational noise. The different distributions behave similarly in this test, slight differences have been found only. The results with STFT (black) and CWD (grey) are displayed on the figure, the curves obtained with PWD are located between the two plotted ones. It is well seen, that the location of the curves are scaled linearly with the amplitude of the added noise. Only very high noise, above the displayed ones, distorts this trend, as the spline interpolation becomes unstable. The amplitude of the observed features of the distributions of the original signal is displayed by a black rectangle in Fig.~\ref{Fig4}. It is clearly demonstrated that even high amount of noise is negligible compared to the real signal in the time-frequency distribution. We have to note that this result depends on the sampling and the signal content of the observations, so it should be repeated for datasets with different characteristics. The nature of noise also influences the above results. We have performed our tests with a broad band (white) noise. For correlated, e.g. color noise, with the same power, false structures appear with higher probability within the frequency range of the noise. Since the power density increases linearly with the inverse of the bandwidth of the noise (with the same total power), similar shift is expected in the $FAP(a)$ curves. However, as our experience suggests, in most applications the effect of a realistic noise turns out to be negligible. 

During the Monte-Carlo simulation the most probable time-frequency distribution due to observational noise can be calculated. For pure Gaussian noise, the expected value of the time-frequency distribution $<T(t,f)>$ is the same for all time-frequency pairs. The averaging and spline smoothing interpolation, however, acts as a low frequency filter, so $<T(t,f)>$ depends on $f$. Because of the uneven sampling this filtering also depends on the time. Therefore, $<T(t,f)>$ of the noise components can be used as a combined time-frequency transfer function of the preprocessing of the data. Fig.~\ref{Fig5} displays this transfer function for the added noise of $\sigma_n=0.04$ for STFT and PWD. With PWD the most probably range of spurious components are narrower in time than that with STFT. This test confirms the above findings, that in the high frequency part of the distribution (around 1 year) the amplitude of the signal is decreased. For times where the sampling is very poor, in addition, an amplification is possible at low frequencies. This amplification is located on the time-frequency plane at periods which are comparable to the length of the longest gaps. However, according to our experience, with a thorough interactive control of the parameters, this effect can be drastically reduced. In the Monte-Carlo test of course it is not possible to do the same fine tuning, and as results the structures are visible in Fig.~\ref{Fig5}.

\section{Application to solar cycles}

Long-term solar records have been  analysed by several authors to date. A review of solar cycle evolution is presented by Usoskin \& Mursula (\cite{usoskin}), with an impressive list of references on the subject. The long-term behaviour of the sunspot group numbers were  analysed using wavelet technique by Frick et al. (\cite{frick}) who plotted the changes of the Schwabe cycle (length and strength) and studied the grand minima. However, there are still remaining features which can be revealed by the use of time-frequency methods. Here we present some of these characteristics from sunspot numbers and solar radio data.

\subsection{Variations Related to the Schwabe Cycle}

  \begin{figure}
   \centering
   \includegraphics[width=8.0cm]{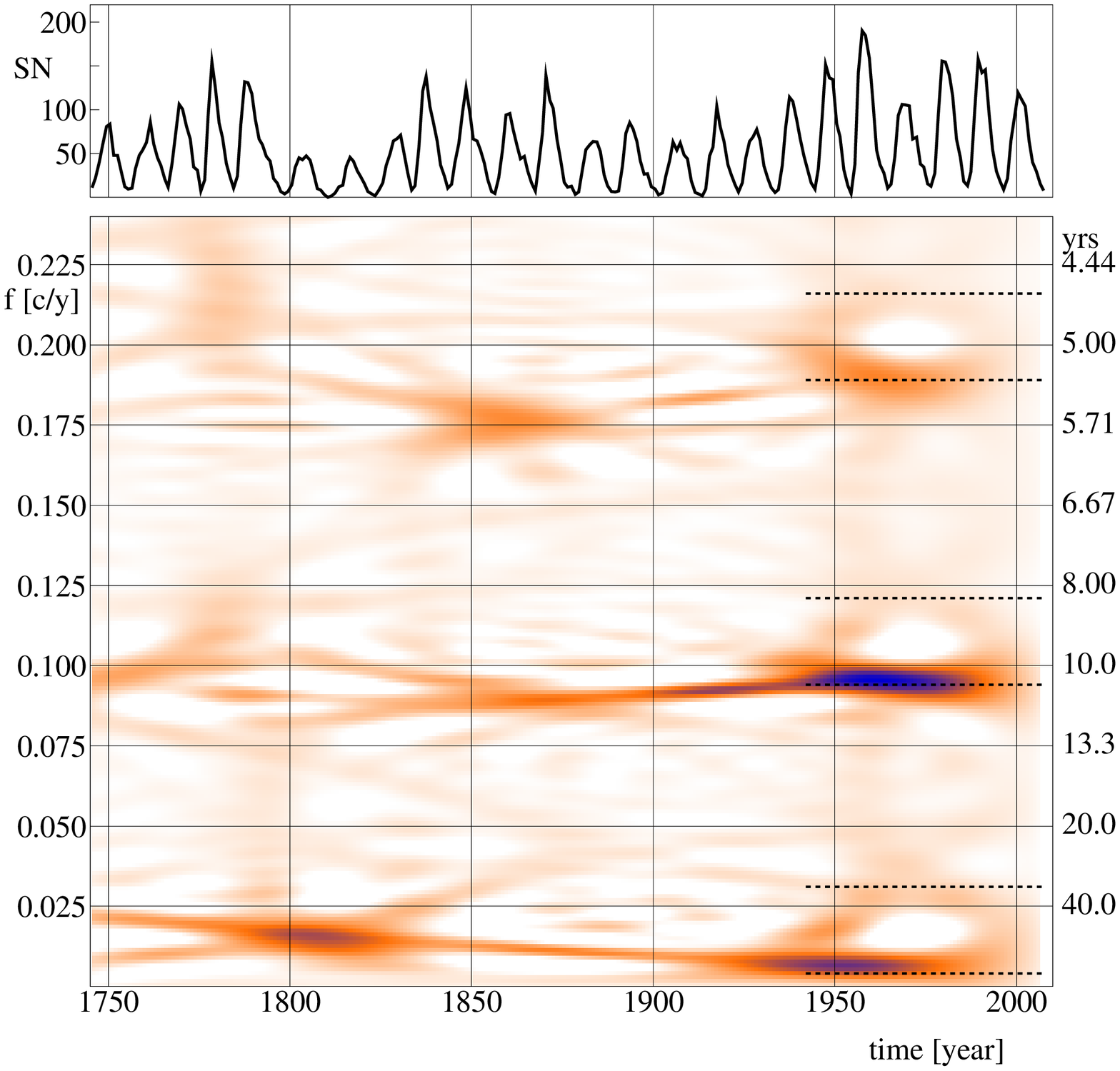}\vspace*{5mm}
   \includegraphics[width=8.0cm]{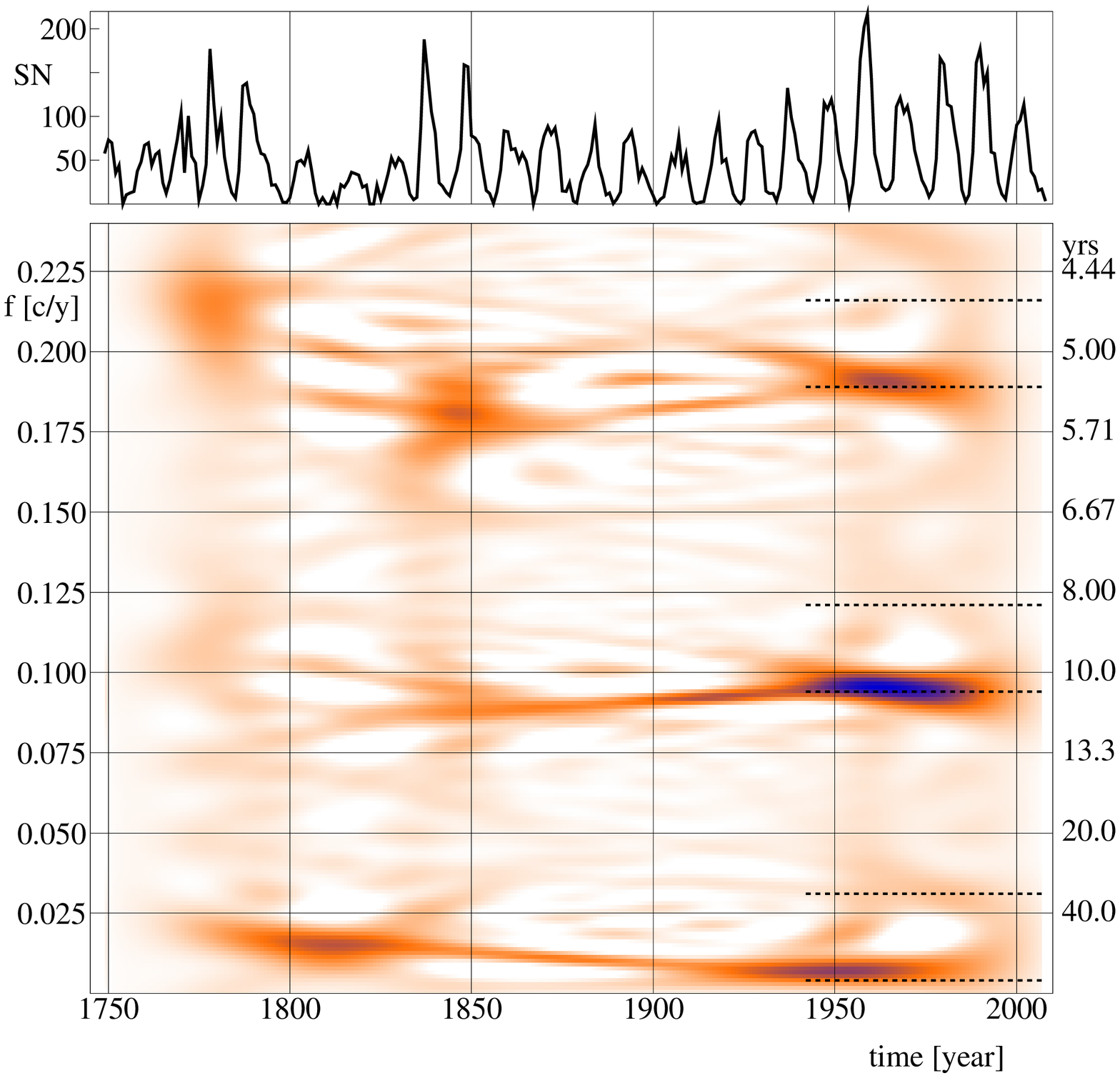}\vspace*{5mm}
   \includegraphics[width=8.0cm]{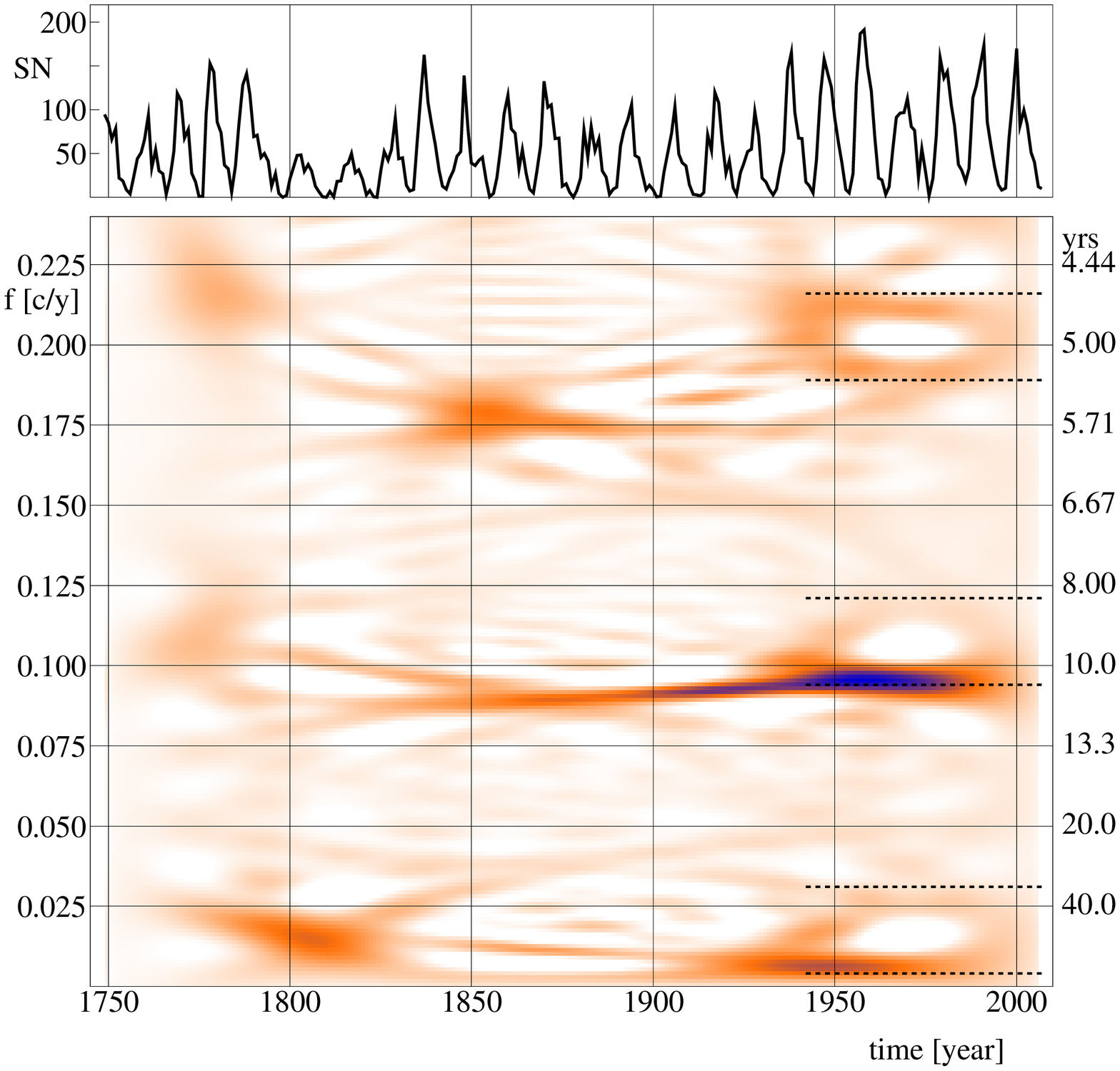}
   
   \caption{ PWD of the yearly sunspot number time series (top) and the same of the datasets generated from the January (middle) and July (bottom) monthly averages 
            }
              \label{Fig6}
    \end{figure}

  \begin{figure}
   \centering
   \includegraphics[width=3.8cm]{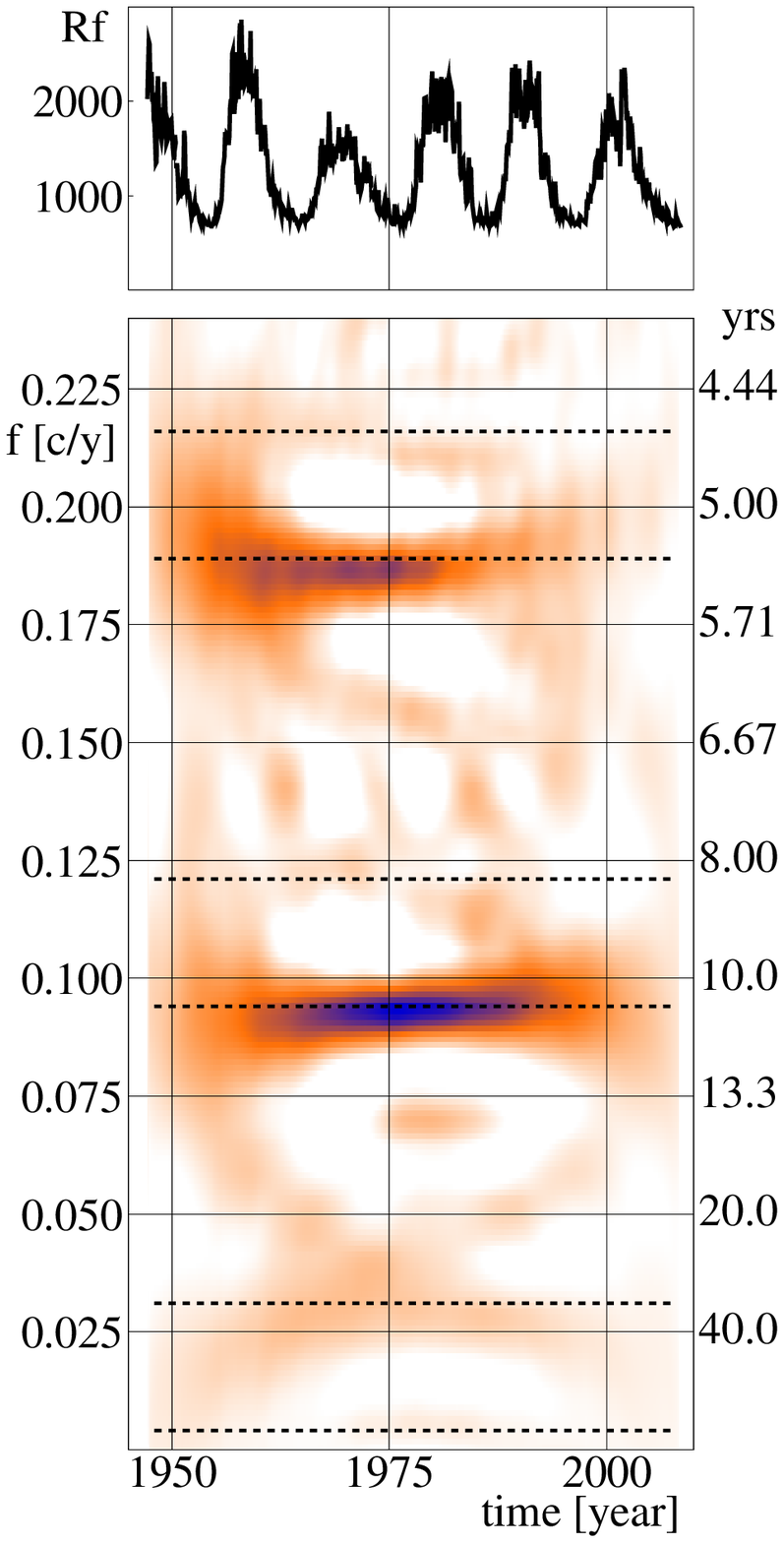}\hspace*{5mm}
   \includegraphics[width=2.3cm]{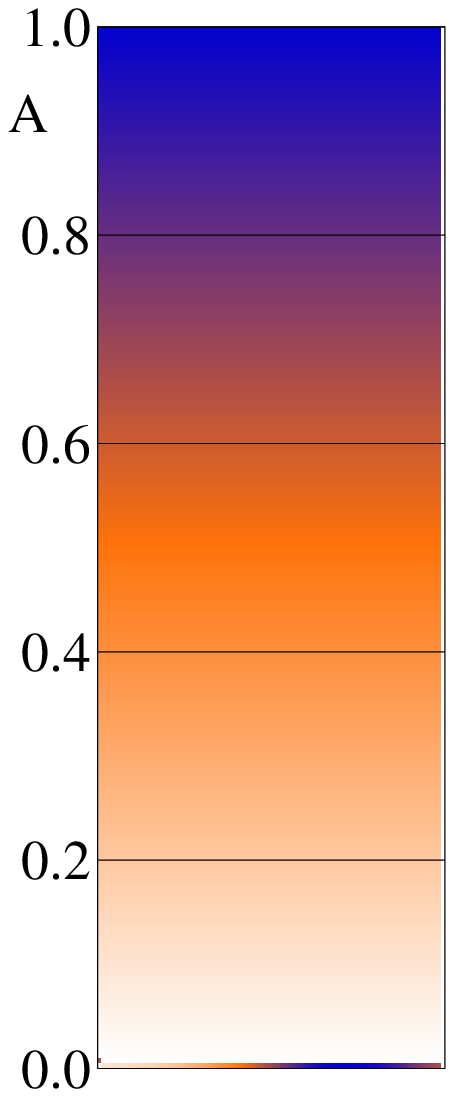}
   \caption{ PWD of the monthly solar radio flux data.
            }
              \label{Fig7}
    \end{figure}

It is well known that the sunspot number is not the best physical indicator of solar activity, however it is the only direct measure we have for a relatively long time-span. Moreover, the sunspot number has an intrinsic fluctuation because of the arbitrariness of the number of visible spots and groups. If we assume that the averages of e.g. the January data are independent measures from that of the July data (i.e., the 'noise' for the averages of a given month are mostly independent from that of the month half a year before and after), then it is possible to construct two independent datasets: one for the January  and one for the July data. The comparison of the results for the two sets gives an idea about the errors in the time-frequency distributions. Fig.~\ref{Fig6} (middle and bottom panels) exhibits the time-frequency distributions for periods longer than 2.5-yr for both data subsets.  The  PWD of the two subsets of the data is hardly distinguishable for periods longer than $\approx 4$ years. 

In Fig.~\ref{Fig6} (top)  we present the  PWD of the whole yearly averaged sunspot data. The STFT (Fig.~\ref{Fig2}, bottom) shows basically the same structure, as the  PWD, but with less visibility and resolution. Also, because the STFT does not contain any false cross terms it validates the higher resolution results of the PWD. 

Both the Schwabe and the Gleissberg cycles can be easily identified in Fig.~\ref{Fig6}. In the followings we denote the instantaneous frequencies of these cycles by $f_S$ and $f_G$, respectively. The modulation of the Schwabe cycle is well known. The period of the main variation of solar activity fluctuates between 9 and 13 years which is well seen in our results (Fig.~\ref{Fig2} and Fig.~\ref{Fig6}). However, the previous investigations neglected to check the harmonics of the frequency $f_S$.  Since the solar cycle is not a sinusoidal variation one should find some power at $2f_S$ and possibly at $3f_S$. Since the harmonic frequencies are synchronised to $f_S$, by the investigation of these harmonics one gets an independent estimate of the variation of the cycle length. The inspection of the history of the $2f_S$ component confirms that around 1780 the cycle had a period of about 9 years, then after the minimal activity it increased to 11 years.

Till the middle of the XXth century the period  decreased to 10  years, then it made a swing to a longer period. It is noticeable that the amplitude of the $2f_S$ component is very small when the Schwabe period changes. It looks as if there were some locked phases, with almost constant period, when the power at $2f_S$ is high.  Another important feature related to the nature of the $2f_S$ component is that its amplitude is increased for the solar cycles with high amplitude more than expected from the ratio of the maximum levels. Since the even harmonics indicate the asymmetry of the form of a solar cycle (symmetric waveforms have only odd harmonics) -- this is another indication of the Waldmeier effect and related phenomena (see e.g. Cameron and Sch\"ussler \cite{cameron}).

The 'bubble' like structures in time-frequency plots indicate that for a finite time interval two frequencies coexist close to each other. The power is clearly split into two frequency parts after 1950, and it is best visible around 1970. The frequency splitting is the same at all three places ($\delta f= 0.025-0.03~c/y$) which indicates a nonlinear connection between different periodicities. The splitting is also visible in the other time-frequency distributions (STFT, CWD), but it is the most prominent in PWD as already mentioned in Section 4.3. As an additional test we have calculated the time-frequency distribution from the 10.7cm radio flux of the Sun, which is available from 1947, and is a good proxy for its activity. The solar radio dataset was recorded at the Dominion Radio Astrophysical Observatory (DRAO) at Penticton, Canada (Tapping \& Charrois \cite{DRAO}), three times a day, and are given in solar flux units or Janskys. The result is plotted in  Fig.~\ref{Fig7}, and it is well seen that the same structure is found in this diagram, than that of Fig.~\ref{Fig6} from sunspot numbers; however, we can not resolve the Gleissberg cycle from the radio data because of the short time base. The possible frequencies appearing in the splitting of the power are indicated by horizontal dotted lines in Fig.~\ref{Fig6} and Fig.~\ref{Fig7}. This test strongly confirms the physical origin of the structure that we have uncovered. To unfold all the regularly spaced frequencies ($f_G$, $f_G+\delta f$, $k*f_S$ and $k*f_S+\delta f$, k=1,2) at least one more frequency is needed in addition to the well known ones ($f_G$ and $f_S$, i.e., the Gleissberg and the Schwabe frequencies, respectively). 

\subsection{Long term variations in the solar cycle}

  \begin{figure*}
   \centering
   \includegraphics[width=15.0cm]{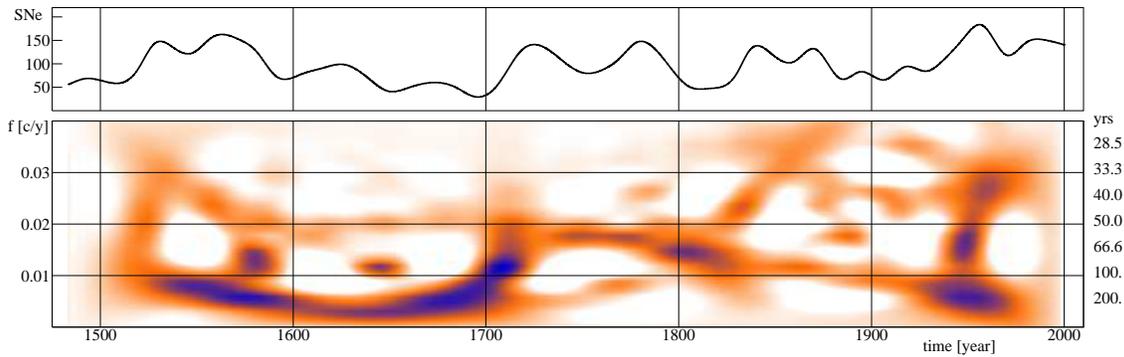}
   \caption{ PWD of the long term envelope of the sunpot data.
            }
              \label{Fig8}
    \end{figure*}

The temporal evolution of the Gleissberg cycle can also be seen on the time-frequency distribution of the solar data. The Gleissberg cycle has been found to be variable as the Schwabe cycle.  It has two higher amplitude occurrences: first around 1800 (during the Dalton minimum), and then around 1950. A very interesting behaviour is the continuous decrease of the frequency (increase of period). While near 1750 the cycle length was about 50-yr it lengthened to approximately 130-yr by 1950.  This period variation explains why different works give different periods of the Gleissberg cycle. This systematic change in the frequency does not support the conclusion of Lawrence, Cadavid and Ruzmaikin (\cite{lawrence}) that the power at long periods is a fingerprint of a period doubling cascade.  The regular pattern of frequency splitting after 1950 gives the possibility to estimate the value of the Gleissberg period for the last decades. $\delta f = 0.03~c/y$ can be precisely determined from the splitting at $f_S$ and $2f_S$. Similarly $f_G+\delta f = 0.031-0.032~c/y$ is well defined. From these two ranges ($\delta f$= 0.025-0.03 and $f_G+\delta f$= 0.031-0.032, so  $f_G < 0.007~c/y$) we suggest that the period of Gleissberg cycle has already reached a level above 140 years. This result agrees with the the rate of frequency decrease from 1800 to 1950. 

We extended the Schove (\cite{schove}) series which lists activity maxima, minima and amplitudes, based mostly on aurora records before 1610, and using optical records afterwards, by the envelope of the recent sunspot data. The dates and values of the maxima were spline smoothed and interpolated to get an equally sampled envelope curve. The last part of the time-frequency distribution of the envelope displays the same structure as the low frequency part of the PWD of the recent data (Fig.~\ref{Fig6}). Fig.~\ref{Fig8} exhibits the smoothed Schove series together with the variation of the Gleissberg cycle for 500 years. 

The Gleissberg cycle had a long period during the Maunder minimum. After 1700 the period jumped to a lower value (50 years) and started a slow increase.  If we accept that the  Gleissberg cycle gives  the  primary indication of the long term behaviour of sunspot numbers, we can interpret the 500-year-long history of solar activity as follows. The long period of the Gleissberg cycle is a good indicator of an extended Maunder minimum, while the subsequent, sudden increase of the Gleissberg period (after 1700) coincides with its termination. An increase of the amplitude of the Gleissberg cycle around 1800, when its period was relatively short, resulted in a shorter (Dalton) minimum. During the last century the Gleissberg period became long, but now indicating a long lasting active stage of the Sun. Can we derive any prediction about the possible future of solar cycles from the extension of this history? The answer depends on the nature of the newly found splitting of the frequencies. If it is a sign of the restart of the Gleissberg cycle at shorter period, then we can expect faster changes in the long term activity level, perhaps a sudden termination of the high activity period. If the 30 year cycle (i.e.,$f_G+\delta f \approx 0.032$, $\sim 30$-yr) is just a temporary one, then a very slow decrease of activity level is expected. To rule out or confirm any of the two possibilities we have to wait for further observations or construct a physical model which is able to produce all the observed behaviour and unfold the nature and connection of all the different periodicities.

Unfortunately, at present we are very far from the construction of a proper physical model of the Sun, if it is at all possible. Bushby and Tobias (\cite{bushby}) discussed the possibilities of predicting the solar cycle through mean-field models, in which the cycle modulations originated from stochastic or deterministic processes. They found that good prediction even of the nearest cycles is impossible in both cases. However, time-series analysis techniques, which find a mapping function for the data, may give better results at least for a short-term prediction.

\section{Conclusions}

   \begin{enumerate}
      \item It is demonstrated that time-frequency distributions are useful tools for analysing nonstationary time series: the short-term Fourier transform. the Wigner distribution and its extensions are more useful than wavelets.
      \item Long-term changes (cycles) of active stars can be correctly identified with the ill-sampled observational data, after careful preprocessing.
      \item We found an extremely complicated multi-scale evolution in the solar activity. All the observed features in the time-frequency history of the Sun should provide strong constraints on modelling the solar magnetism.
   \end{enumerate}

Whether it is possible to predict the long scale future of solar activity from the similarities of the structures  that have emerged in the last decades and those in the early history of the Sun, is uncertain. Its verification would be an important future project both in theory and data analysis, because of the probable connection between the long term changes in our climate and solar activity (see Velasco \& Mendoza (\cite{velasco_mendoza}) and references therein, for more).

\begin{acknowledgements}
We  are grateful to A.F. Lanza and W. Soon for useful discussions during the preparation of the manuscript. Our thanks are due to the referee, Dr. F. Baudin, whose suggestions helped to improve this paper as well as the companion paper. K.O. acknowledges supports from the Hungarian Research Grants OTKA T-048961 and T-068626. 
\end{acknowledgements}

\end{document}